\documentclass[
showpacs,preprintnumbers,
nofootinbib,
 amsmath,amssymb,
 aps,
prd,
twocolumn
]{revtex4-1}

\usepackage[utf8]{inputenc}
\usepackage{epsf}
\usepackage{float}
\usepackage{graphicx} 
\usepackage{amssymb}
\usepackage{dcolumn}
\usepackage{amsmath}
\usepackage{hyperref}
\usepackage{multirow}
\usepackage{booktabs}
\usepackage{siunitx}
\usepackage{color}

\hypersetup{colorlinks=true,linkcolor=blue,citecolor=blue,filecolor=blue,urlcolor=blue}

\begin{document}

\title{Constraining neutrino mass with tomographic weak lensing \\one-point probability distribution function and power spectrum}

\author{Jia Liu}
\thanks{NSF Astronomy and Astrophysics Postdoctoral Fellow}
\email{jia@astro.princeton.edu}

\author{Mathew S. Madhavacheril}

\affiliation{ {Department of Astrophysical Sciences, Princeton University, Princeton, NJ 08544, USA}} 

\newcommand{\msm}[1]{\textcolor{magenta}{(MM: #1)}}                                                                                                  

\date{\today}

\begin{abstract}
We study the constraints on neutrino mass sum ($\Sigma m_\nu$) from the one-point probability distribution function (PDF) and power spectrum of weak lensing measurements for an LSST-like survey, using the \texttt{MassiveNuS} simulations. The PDF provides access to non-Gaussian information beyond the power spectrum. It is particularly sensitive to nonlinear growth on small scales, where massive neutrinos also have the largest effect. We find that tomography helps improve the constraint on $\Sigma m_\nu$ by 14\% and 32\% for the power spectrum and the PDF, respectively, compared to a single redshift bin. The PDF alone outperforms the power spectrum in constraining $\Sigma m_\nu$. When the two statistics are combined, the constraint is further tightened by 35\%. We conclude that weak lensing PDF is complementary to the power spectrum and has the potential to become a powerful tool for constraining neutrino mass.
\end{abstract}

\maketitle

\section{Introduction}\label{sec:intro}

The sum of neutrino masses ($\Sigma m_\nu$) is now known to be at least 0.06~eV, after the discovery of oscillations between their flavor eigenstates~\cite{Becker-Szendy1992,Fukuda1998,Ahmed2004}. Cosmic neutrinos affect the expansion history and growth of structure in the Universe, and hence observations of large-scale structure can be used to constrain $\Sigma m_\nu$~(see reviews by~\cite{LesgourguesPastor2006,Wong2011}).
At present, the tightest bound on $\Sigma m_\nu\leq0.12$~eV comes from the 2018 \texttt{Planck} analysis, combining cosmic microwave background (CMB) temperature and polarization, CMB lensing, and baryon acoustic oscillation (BAO) measurements~\cite{planck2018}. Measuring the value of $\Sigma m_\nu$ is one of the key science goals of next generation galaxy surveys such as the LSST\footnote{Large Synoptic Survey Telescope: \url{http://www.lsst.org}}~\cite{LSSTscienceBook}, WFIRST\footnote{Wide-Field Infrared Survey Telescope: \url{http://wfirst.gsfc.nasa.gov} }, and Euclid\footnote{Euclid: \url{http://sci.esa.int/euclid} }and CMB surveys such as the Simons Observatory\footnote{Simons Observatory: \url{https://simonsobservatory.org}}~\cite{SO2018} and CMB-S4\footnote{CMB-S4: \url{https://cmb-s4.org/}}~\cite{CMB-S42016}. 

Weak gravitational lensing by the large-scale structure is a promising tool for precision cosmology~(see a recent review by~\cite{Kilbinger2015}). Photons emitted from distant galaxies are deflected by the intervening matter---be it baryonic or cold dark matter (CDM). Lensed galaxies are (de)magnified in brightness and distorted from their intrinsic shape. From statistical measurements of galaxy shapes, we can infer the matter distribution between us and the lenses. Furthermore, by splitting background galaxies into several redshift bins, i.e. the ``redshift tomography'' technique, we can gain insights into the evolution of structure growth. Statistical measurements of weak lensing have been achieved in the past decade and are now commonly used for constraining cosmology~\cite[e.g.][]{Schrabback2010,Heymans2012,Hildebrandt2017,Mandelbaum2017,2017DES}.

In this work, we study the information stored in weak lensing one-point probability distribution function~(PDF). Comparing to the commonly used Gaussian (or second-order) statistics---the two-point correlation function and its Fourier transformation, the power spectrum---PDF can capture additional non-Gaussian (or higher-order) information. The origin of non-Gaussianity in the lensing field is the nonlinear growth of structure, which is more prominent at small scales and at late times. Non-Gaussian statistics have been tested both theoretically and on data, and are found to be powerful in improving cosmological constraints\footnote{For example, higher order moments~\cite{Bernardeau1997,Hui1999,vanWaerbeke2001, Takada2002,Zaldarriaga2003,Kilbinger2005,Petri2015,Peel2018}, 
three-point functions~\cite{Takada2003,Vafaei2010,Fu2014}, 
bispectra~\cite{Takada2004,DZ05,Sefusatti2006,Berge2010}, 
peak counts~\cite{Jain2000b,Marian2009,Maturi2010,Yang2011,Marian+2013,Liu2015,Liu2014b,Lin&Kilbinger2015a,Lin&Kilbinger2015b,Kacprzak2016,Peel2018}, 
Minkowski functionals~\cite{Kratochvil2012,Shirasakiyoshida2014,Petri2013,Petri2015}.}, compared to using Gaussian statistics alone.

Non-Gaussian statistics are particularly interesting for constraining $\Sigma m_\nu$, because they are most powerful on small scales, where massive neutrinos also leave the strongest signature. With large thermal velocities, cosmic neutrinos stream out of CDM potential wells freely, suppressing the growth of structure below the ``free-streaming scale''.  For neutrino masses within the current constraints, the free-streaming scale is around 100 Mpc.  
Ref.~\cite{Liu2016pdf} studied the PDF of CMB lensing for a CMB-S4 like survey, and found only mild improvement on $\Omega_m$ and $\sigma_8$ from the power spectrum constraint, because CMB lensing probes structure at high redshift where growth is mostly linear. We expect the PDF to be more powerful for galaxy weak lensing, as nonlinear structures are more prominent at low redshift. Fisher-matrix based forecasts by Ref.~\cite{Patton2017} using both the weak lensing power spectrum and the PDF for a single redshift bin  showed a factor of 2--3 improvement on $\Omega_m$ and $\sigma_8$ from that of the power spectrum alone. Measurements of the thermal Sunyaev-Zel'dovich one-point probability distribution function have also been shown to have cosmological sensitivity~\cite{Hill2014tSZ1pt,Planck2016tSZ}.

The goal of this paper is to forecast the constraints on $\Sigma m_\nu$ from the weak lensing PDF and power spectrum, for an LSST-like survey, using the Cosmological Massive Neutrino Simulations~(\texttt{MassiveNuS}). Our work is only the first step to explore the power of non-Gaussian statistics to constrain neutrino mass. At relevant scales in this work---well into the so-called ``one-halo regime'' where the internal structure of halos are probed---baryonic feedback is also relevant. Modeling baryonic effects, so far mainly done at the power spectrum level~\cite{Semboloni2011,Huang2018,Parimbelli2018}, will be an important next step to take for higher-order statistics.

The paper is organized as follow. First, we describe our simulations, statistical measurements, and likelihood analysis, in section~\ref{sec:tech}. We show results in \ref{sec:results}, including the effect of massive neutrinos on the PDF, the power of tomography using multiple redshift bins, and joint constraints of the power spectrum and the PDF. Finally, we conclude in section~\ref{sec:conclusion}. 

\section{Analysis}\label{sec:tech}

\subsection{Simulations}

We use mock lensing maps from the Cosmological Massive Neutrino Simulations~(\texttt{MassiveNuS})~\cite{Liu2018MassiveNuS:Simulations}\footnote{The \texttt{MassiveNuS} data products, including galaxy and CMB lensing convergence maps, N-body snapshots, halo catalogues, and merger trees, are publicly available at \url{http://ColumbiaLensing.org}.}. Here we briefly introduce the simulations, and refer the reader to \cite{Liu2018MassiveNuS:Simulations} for more detailed descriptions and code validation.

\texttt{MassiveNuS} consists of a suite of 101  flat-$\Lambda$CDM N-body simulations, with three varied parameters: the neutrino mass sum $\Sigma m_\nu$, the total matter density $\Omega_m$, and the primordial power spectrum amplitude $A_s$.  They cover the range $\Sigma m_\nu$=[0, 0.62]~eV, $\Omega_m$=[0.18, 0.42], $A_s\times 10^9$=[1.29, 2.91]. The simulations use the public code \texttt{Gadget-2}~\cite{springel2005}, with a box size of 512~Mpc$h^{-1}$ and 1024$^3$ CDM particles, accurately capturing structure growth at $k<$10~$h$~Mpc$^{-1}$. Massive neutrinos are treated using linear perturbation theory and their clustering is sourced by the full nonlinear matter density. The neutrino patch code~\citep{AB2013,Bird2018} has been tested robustly against particle neutrino simulations, and the total matter power spectrum is found to agree with theory to within 0.2\% for $\Sigma m_\nu < 0.6$ eV. 

Weak lensing convergence~($\kappa$) maps are generated for five source redshifts $z_s$=0.5, 1.0, 1.5, 2.0, 2.5, using the ray-tracing code \texttt{LensTools}~\cite{Petri2016Lenstools}\footnote{\url{https://pypi.python.org/pypi/lenstools/}}. For each cosmological model and source redshift, 10,000 map realizations are generated. All maps are 512$^2$~pixels and 3.5$^2$=12.25~deg$^2$ in size. For each realization, the maps at different source redshifts are ray-traced through the same large-scale structure and hence are properly correlated. 

To create LSST-like mocks, we follow the estimation in LSST Science Book~(section~3.7.2 of \cite{LSSTscienceBook})~\footnote{We note that defining the LSST survey parameters is still on-going work and is science-dependent, also see Ref.~\cite{DESC2018}.}. We assume the total galaxy number density $n_{\rm gal}$=50~arcmin$^{-2}$ with source redshift distribution,
\begin{align}\label{eq:Pz}
n(z) &\propto z^\alpha \exp[-(z/z^*)^\beta]
\end{align}
where $\alpha$=2, $z^*$=0.5, $\beta$=1. Assuming $\Delta z_s$=0.5 for each source redshift bin, we obtain the number density for each source redshift.
\begin{center}
\begin{tabular}{c|c|c|c|c|c} 
\hline
$z_s$	&	0.5	&	1.0	&	1.5	&	2.0	&	2.5	\\
\hline
$n_{\rm gal}$ (arcmin$^{-2}$)	&	8.83 		&	13.25	&	11.15	&	7.36	&	4.26\\
\hline
\end{tabular}
\end{center}
We obtain a smaller total number density of 44.85~arcmin$^{-2}$, as the result of discarding galaxies at $z_s<$0.25 and $z_s>$2.75.
To add galaxy noise to the noiseless $\kappa$ maps, we add to each pixel a number randomly drawn from a Gaussian distribution centered at zero with variance=$\sigma_\lambda^2/(n_{\rm gal}\Omega_{\rm pix})$, where $\sigma_\lambda$=0.3 is the shape noise and $\Omega_{\rm pix}$ is the solid angle of a pixel in unit of arcmin$^2$.

\subsection{Power spectrum and PDF}

We compute the power spectrum and PDF for all  101 $\times$ 5 $\times$ 10,000 mocks. For the power spectrum, we square the Fourier transformation of the map, and compute the average power within each of the 20 linear bins between $\ell_{\rm min}$=100 and $\ell_{\rm max}$=5,000. Overall, there are 15 possible combinations for the power spectrum, from the five redshift bins (five auto-correlations and 10 cross-correlations). Here we use only the five auto-correlations, as we found that these are sufficient to recover most of the Gaussian information.


For the PDF, we first smooth the maps to reduce large contributions from noise. For a strict comparison with the power spectrum, we filter the maps in Fourier space with all modes larger than $\ell_{\rm max}$=5,000 set to 0, and then inverse Fourier transform back to real space. In real space, $\ell_{\rm max}$=5,000 is equivalent to $\approx$2~arcmin. We compute the PDF in each $\kappa$ bin, for 20 linear bins between [$-$3$\sigma_\kappa$, 5$\sigma_\kappa$], where $\sigma_\kappa$ is the standard deviation of the maps at our massless fiducial model with $\Sigma m_\nu$=0.0~eV, $A_s$=$2.1\times10^{-9}$, and $\Omega_m=0.3$. $\sigma_\kappa^{\rm noiseless}$=[ 0.008,  0.016,  0.023,  0.029,  0.034] and $\sigma_\kappa^{\rm noisy}$=[0.041, 0.037, 0.042, 0.053, 0.066] for $z_s$=[0.5, 1.0, 1.5, 2.0, 2.5], respectively.

\subsection{Likelihood}

We forecast the constraints for the power spectrum and PDF separately and jointly. We set our fiducial model to be $\Sigma m_\nu$=0.1~eV, $A_s$=$2.1\times10^{-9}$, and $\Omega_m=0.3$. Here we describe the three critical components of our likelihood analysis: the emulator, the covariance matrix, and the likelihood function. 

We build an emulator for each statistic, which allows us to generate a model power spectrum or PDF at any parameter point. We use the Gaussian Process module implemented in the \texttt{scikit-learn}\footnote{\url{http://scikit-learn.org}} Python package. It takes in the average power spectra or PDFs (over 10,000 realizations) for all models as observations, and interpolates through their cosmological parameters ( $\Sigma m_\nu$, $A_s$, $\Omega_m$). We test the Gaussian Process interpolator by comparing the prediction of a target model (using an emulator built without the model) to the ground truth (i.e. the actual value from the simulation), for 10 models near the massive fiducial model. We find that the interpolator performs well for both statistics, with sub-percent differences and are always within the statistical error (scaled to the LSST sky coverage). 

\begin{figure*}
\begin{center}
\includegraphics[width=0.45\textwidth]{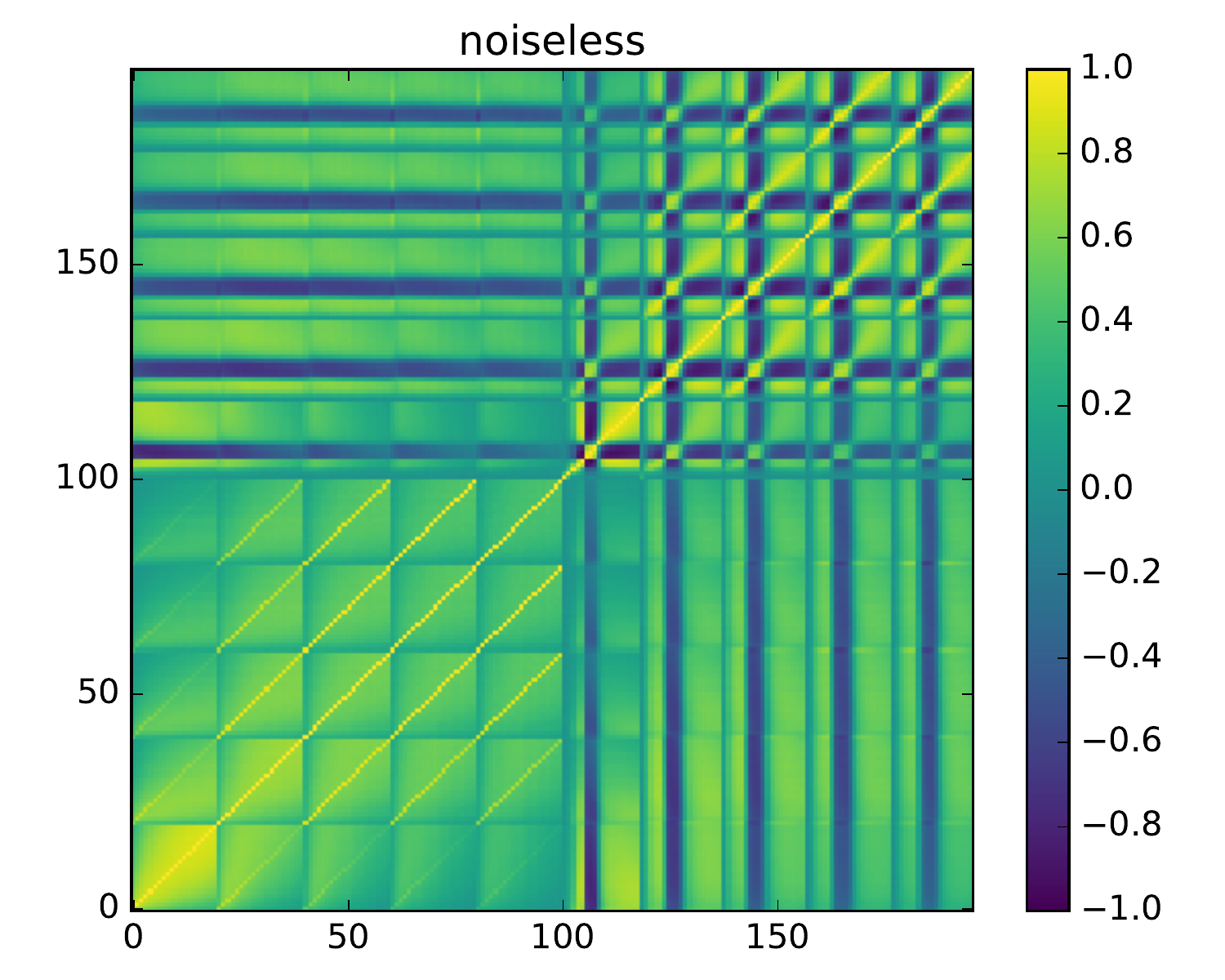}
\includegraphics[width=0.45\textwidth]{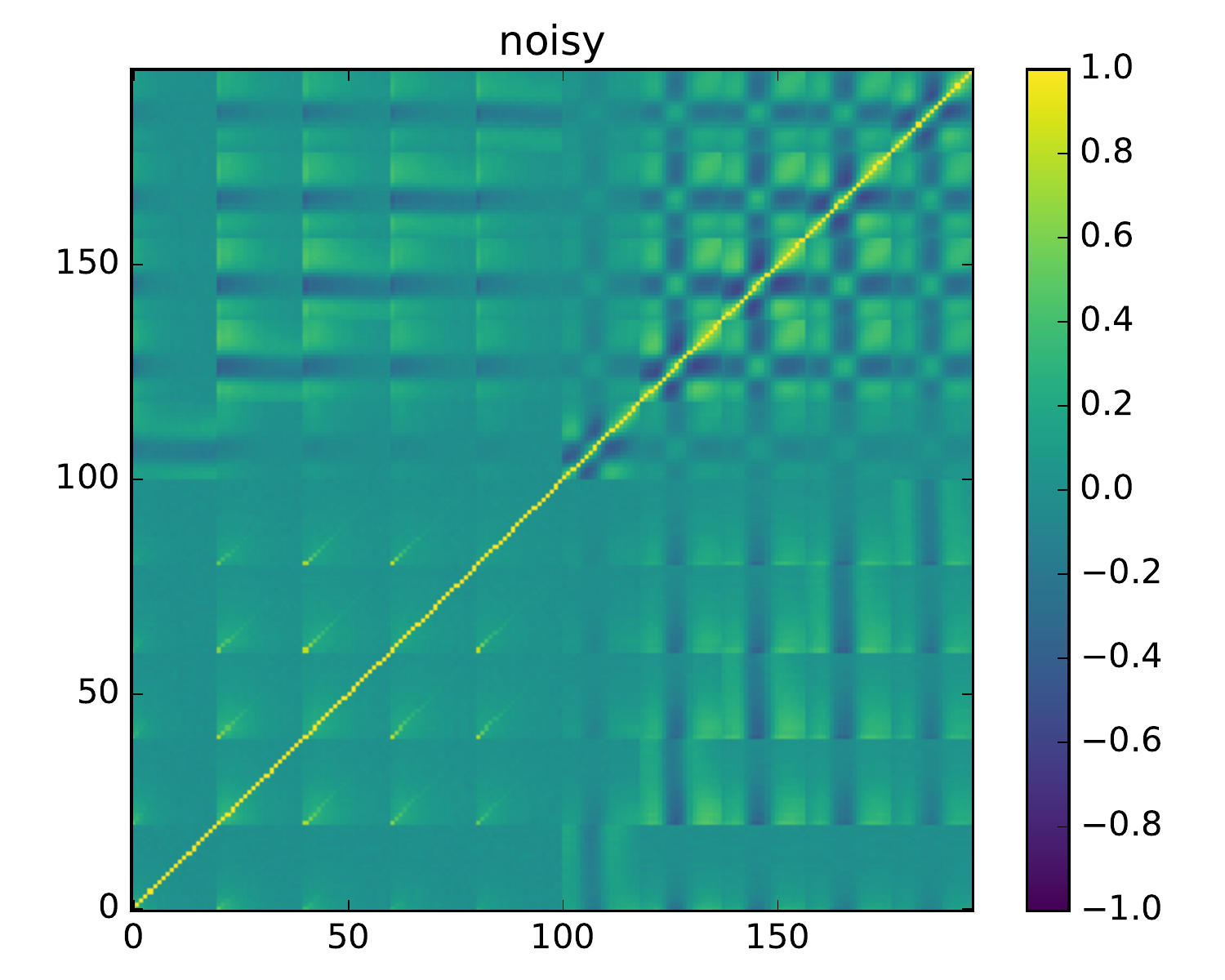}
\end{center}
\caption{\label{fig:cov} Noiseless (left) and noisy (right) covariance matrices, normalized by the diagonal terms. The first 100 bins are the power spectrum bins, and the rest are the PDF bins. Each of the two blocks have five sub-blocks, 
representing the tomographic redshift bins.}
\end{figure*}

To model the covariance matrices, we use an independent set of simulations at the fiducial model, to avoid the correlation between the noise in the emulator and the covariance. We show the covariance matrices for both the noiseless and noisy maps in Fig.~\ref{fig:cov}. In the noiseless case, the power spectrum block (bottom-left 100 bins) shows the usual diagonal behavior, with large off-diagonal terms only at the lowest redshift bin $z_s=0.5$. In contrast, the PDF block (top-right 100 bins) has a complicated check pattern, showing that PDF bins are highly (anti)correlated. In the noisy case, the off-diagonal terms are less prominent, though remain visible. We apply a correction factor $f$=($N_{\rm sim}-N_{\rm bin}-2$)/($N_{\rm sim}-1$) to the inverse covariance ${\bf C}^{-1}$ to account for the limited number of simulations, where $N_{\rm sim}$=10,000 is the number of mock realizations and $N_{\rm bin}$ is the number of bins~\cite{hartlap2007}. We multiply the covariance by the ratio of our map size (12.25~deg$^2$) to the LSST sky coverage (20,000~deg$^2$). 

We assume a Gaussian likelihood for both statistics, with the log likelihood,
\begin{align}
{\bf L} ({\boldsymbol d}|{\boldsymbol p})=-\frac{1}{2}\left({\boldsymbol d}-{\boldsymbol \mu}\right)^T {\bf C}^{-1}\left({\boldsymbol d}-{\boldsymbol \mu}\right)
\end{align}
where ${\boldsymbol d}$ is the ``observation'' vector (in our case, the average statistics at the fiducial model), ${\boldsymbol p}$ is the three parameters in interest, and ${\boldsymbol \mu}$ is the emulator prediction. 
We use the \texttt{emcee}\cite{foremanmackey2013} Python package to implement the Markov chain Monte Carlo (MCMC). We apply a wide flat prior for all parameters, and set ${\bf L}=-\inf$ for $\Sigma m_\nu<0$, i.e. force the neutrino mass sum to be non-zero. We run 1.6 million chains, and discard the first 25\% as burn-in. We have tested that our results are well converged with just 0.3 million chains. We also tested that our results are immune to the initial walker position (a very wide prior vs. a tight ball around the fiducial model).

\section{Results}\label{sec:results}

\subsection{PDF with massive neutrinos}

\begin{figure*}
\begin{center}
\includegraphics[width=0.48\textwidth]{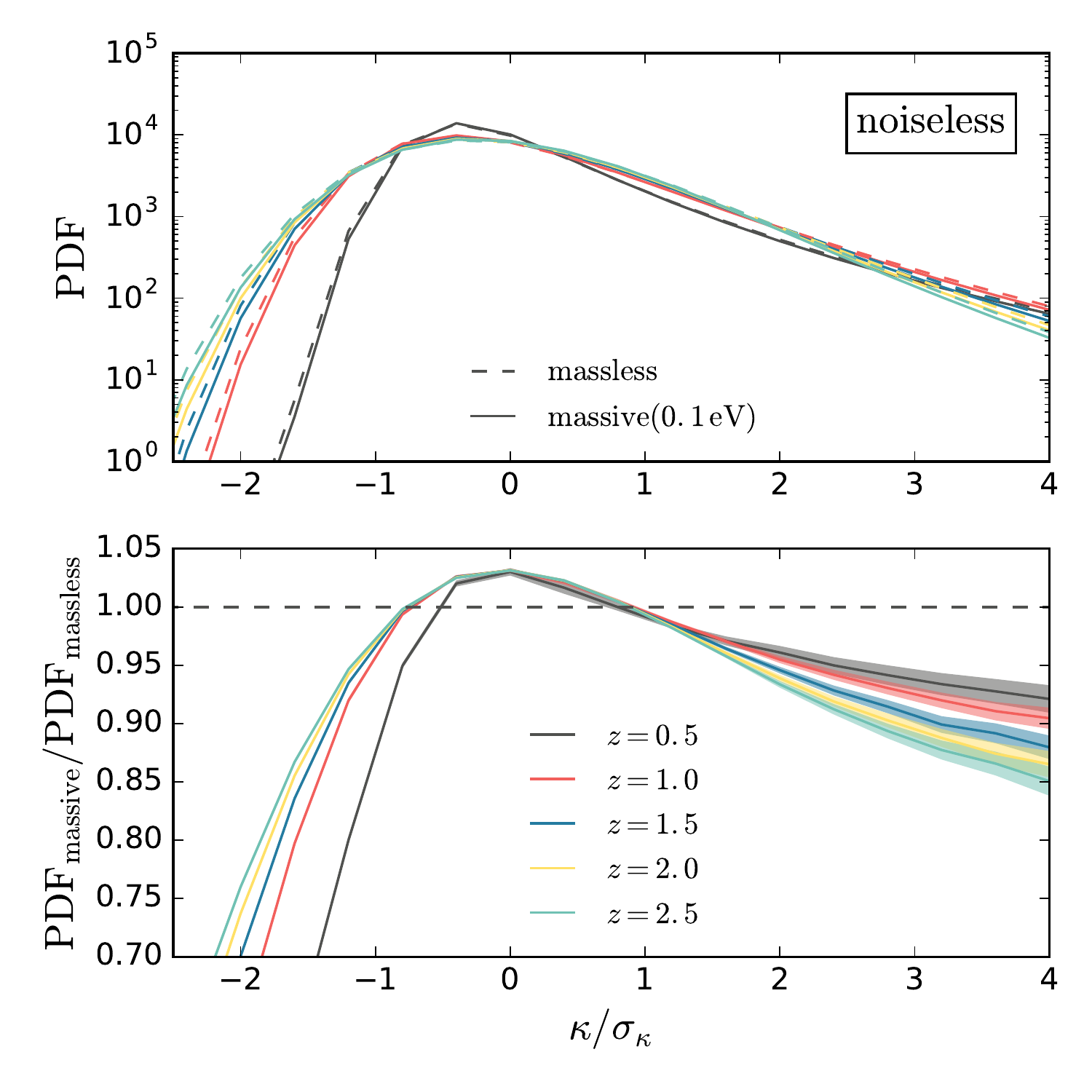}
\includegraphics[width=0.48\textwidth]{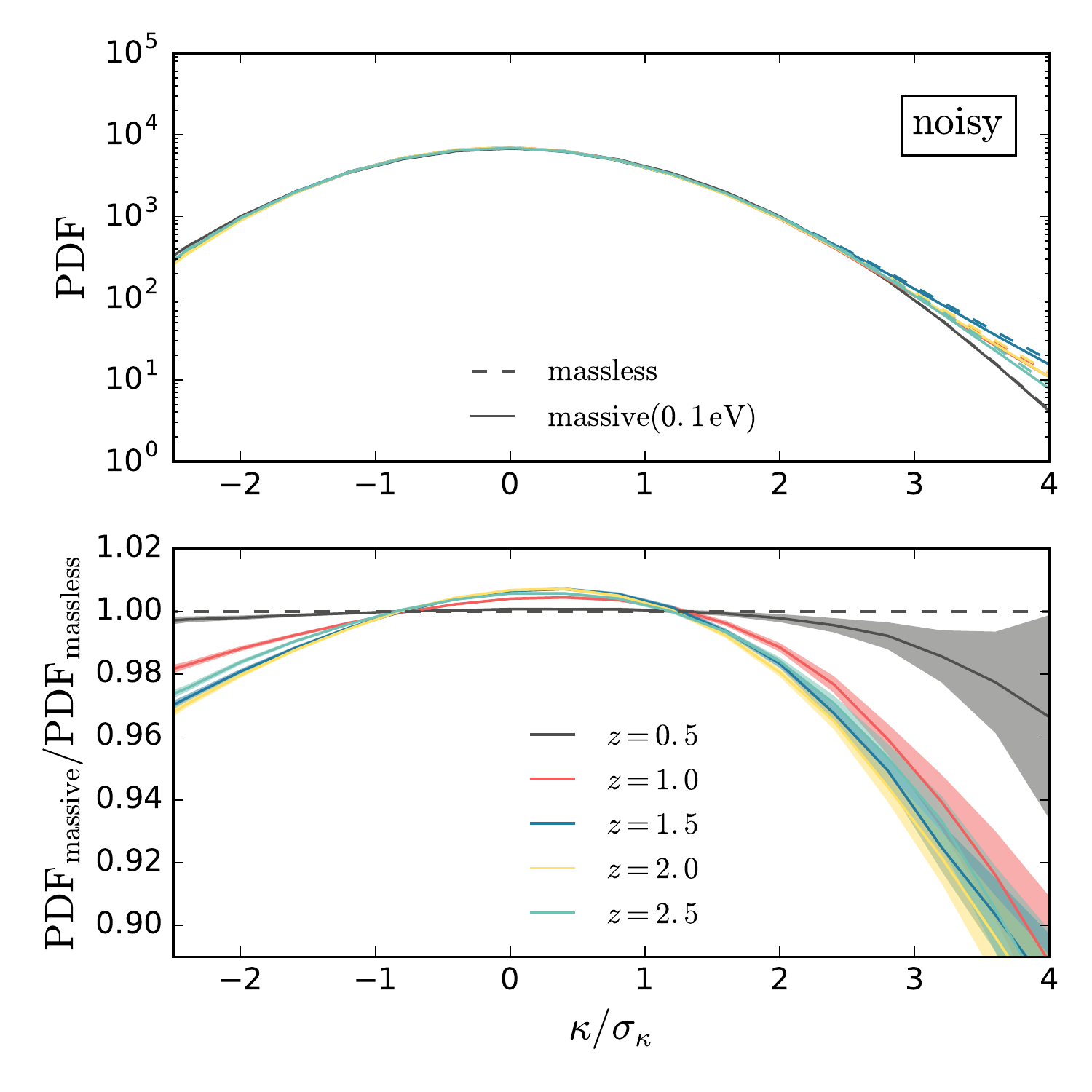}
\end{center}
\caption{\label{fig:pdf} The noiseless (left) and noisy (right) PDFs for the fiducial models, as a function of $\kappa/\sigma_\kappa$, the convergence normalized by the standard deviation for the massless model, where  
$\sigma_\kappa^{\rm noiseless}$=[ 0.008,  0.016,  0.023,  0.029,  0.034], 
$\sigma_\kappa^{\rm noisy}$=[0.041, 0.037, 0.042, 0.053, 0.066] for $z_s$=[0.5, 1.0, 1.5, 2.0, 2.5], respectively. The 95\% confidence level errors, scaled to LSST sky coverage, are shown as colored bands in the lower panels (as they would be invisible in the upper panel due to their small sizes), though we note the bins are highly correlated, as shown in Fig.~\ref{fig:cov}.}
\end{figure*}

We show comparisons of massive and massless neutrino models in Fig.~\ref{fig:pdf}. We first examine the noiseless case (left panels), where the physical effect of massive neutrinos is more transparent. In the upper panel, the PDFs of all tomographic bins show a non-zero skewness, with a long tail at the high $\kappa$ side. This is a clear signature of non-Gaussianity, hinting at additional information beyond the power spectrum. The skewness is larger at lower redshift, due to increasingly nonlinear growth. In the lower panel, we show the ratio between the massive~(0.1~eV) and massless neutrino models, while holding other two parameters fixed. Massive neutrinos suppress both the positive and negative tails of the PDFs---in other words, massive neutrinos result in smaller number of massive halos and troughs (projection of voids along the line of sight). This is not surprising, as we expect the growth of halos and voids to be correlated--- matter falling into CDM potential wells would in turn leave other regions emptier  (though the effect of massive neutrinos on voids can be complicated, see~\cite{Kreisch2018}).  

After we add galaxy noise (right panels), the PDFs (upper panel) become more Gaussian, though the high $\kappa$ non-Gaussian tails remain visible. In the lower panel, the differences between the massive and massless neutrino models are reduced, especially for the low $\kappa$ bins.

\subsection{The power of tomography}

\begin{figure*}
\begin{center}
\includegraphics[width=0.75\textwidth]{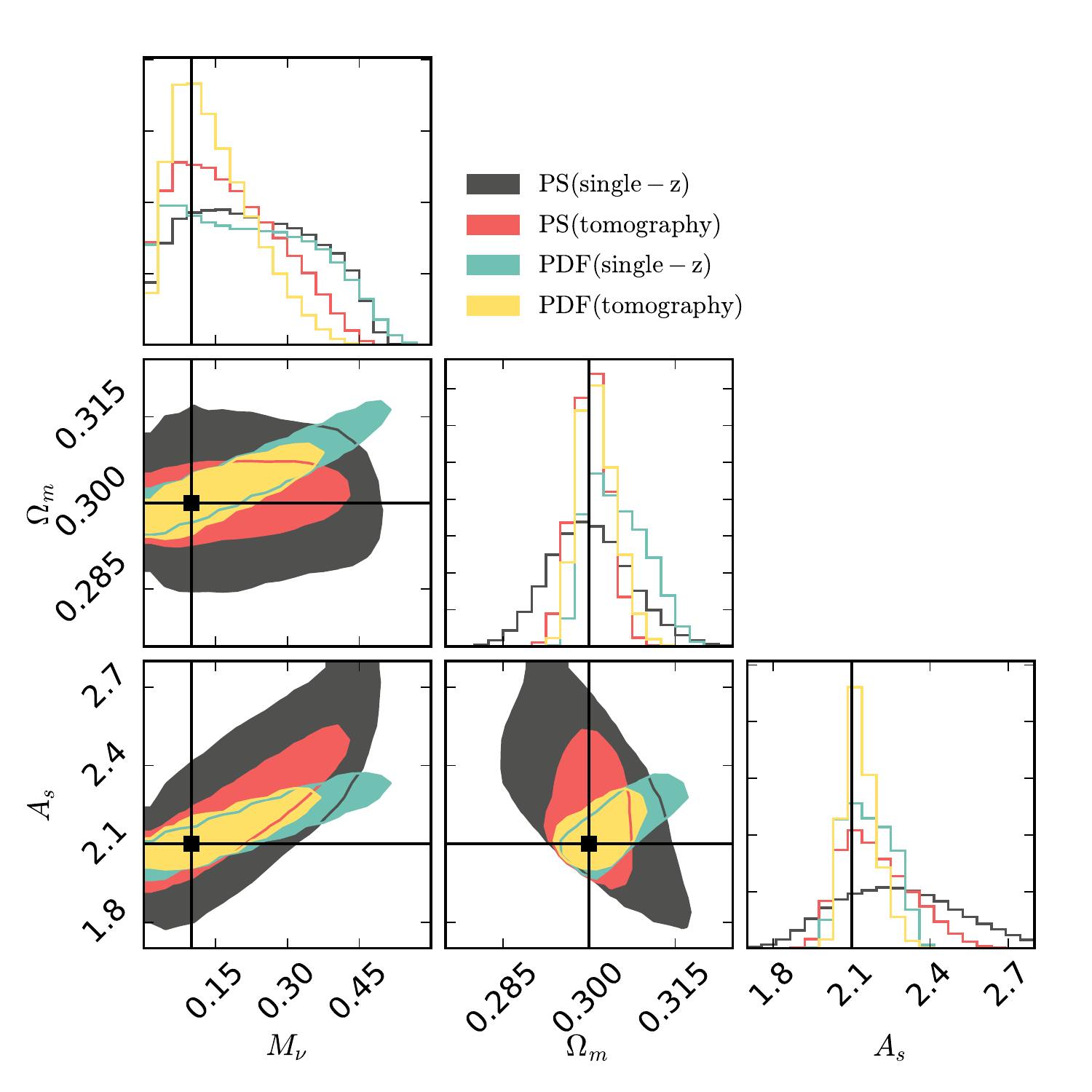}
\end{center}
\caption{\label{fig:tomo} 95\% CL contours for single redshift bin vs. five tomographic bins, for the weak lensing power spectrum~(PS) and PDF. We assume an LSST-like survey. }
\end{figure*}

Weak lensing tomography, i.e. splitting the source galaxies by their redshift, has been proposed as a tool to recover the three-dimensional density field from the two-dimensional projected maps~\cite{Hu1999,Hu2002b,JainTaylor2003,TakadaJain2004c,Hannestad2006}. Tomography has shown to have the potential to tighten the cosmological constraints by up to an order of magnitude\cite{Hu1999,Hannestad2006,Petri2016tomo}, compared to single redshift maps. However, when implemented on data~(see recent measurements of tomographic power spectrum~\cite{heymans2013,Benjamin2013,Kohlinger2016,Hildebrandt2017,Kohlinger2017}), the relative improvement can degrade due to systematics, in particular the uncertainties and biases in the photometric redshift measurements~\cite{Ma2006,Hearin2010}.

To study the relative improvement from tomography, we compute the power spectrum and PDF for single redshift maps, which we created using $\kappa$ maps only at $z_s=1$, the peak of the redshift distribution~(eq.~\ref{eq:Pz}), but with galaxy density $n_{\rm gal}=44.85$~arcmin$^{-2}$, equivalent to the sum of all galaxies in the five tomographic bins.  

We show the 95\%~CL contours for tomography vs. single-$z$ in Fig.~\ref{fig:tomo}. We quantify the improvement using tomography as $\sigma_p^{\rm tomo}/\sigma_p^{\rm single-z}$, where $\sigma_p$ is the two-sided 95\% CL error i.e. the sum of the positive and negative error sizes, on parameter $p$:
\begin{center}
\begin{tabular}{c|c|c|c} 
 \multicolumn{4}{r}{$\sigma_p^{\rm tomo}/\sigma_p^{\rm single-z}$}\\
\hline
$p$	&	$\Sigma m_\nu$  & $\Omega_m$& $A_s$	\\
\hline
power spectrum &	0.86	&	0.42	&	0.37\\
\hline
PDF 			&	0.68	&	0.71	&	0.80\\
\hline
\end{tabular}
\end{center}
For $\Omega_m$ and $A_s$, the improvement is more significant for the power spectrum. In particular, the error sizes from tomography are less than half of that from single-$z$ for $\Omega_m$ and  $A_s$. We also see modest improvement for the PDF, by 20--30\%. For $\Sigma m_\nu$, the PDF error is reduced by 32\%, compared with only 14\% for the power spectrum error.

%
%


In the case of power spectrum, we only include the five auto-correlations, omitting the other possible 10 cross-correlations between different redshift bins. We find that including the cross powers only adds marginal percent level improvement to the auto powers, and hence decide to discard them for simplicity. 

\subsection{Joint likelihood}
\begin{figure*}
\begin{center}
\includegraphics[width=0.75\textwidth]{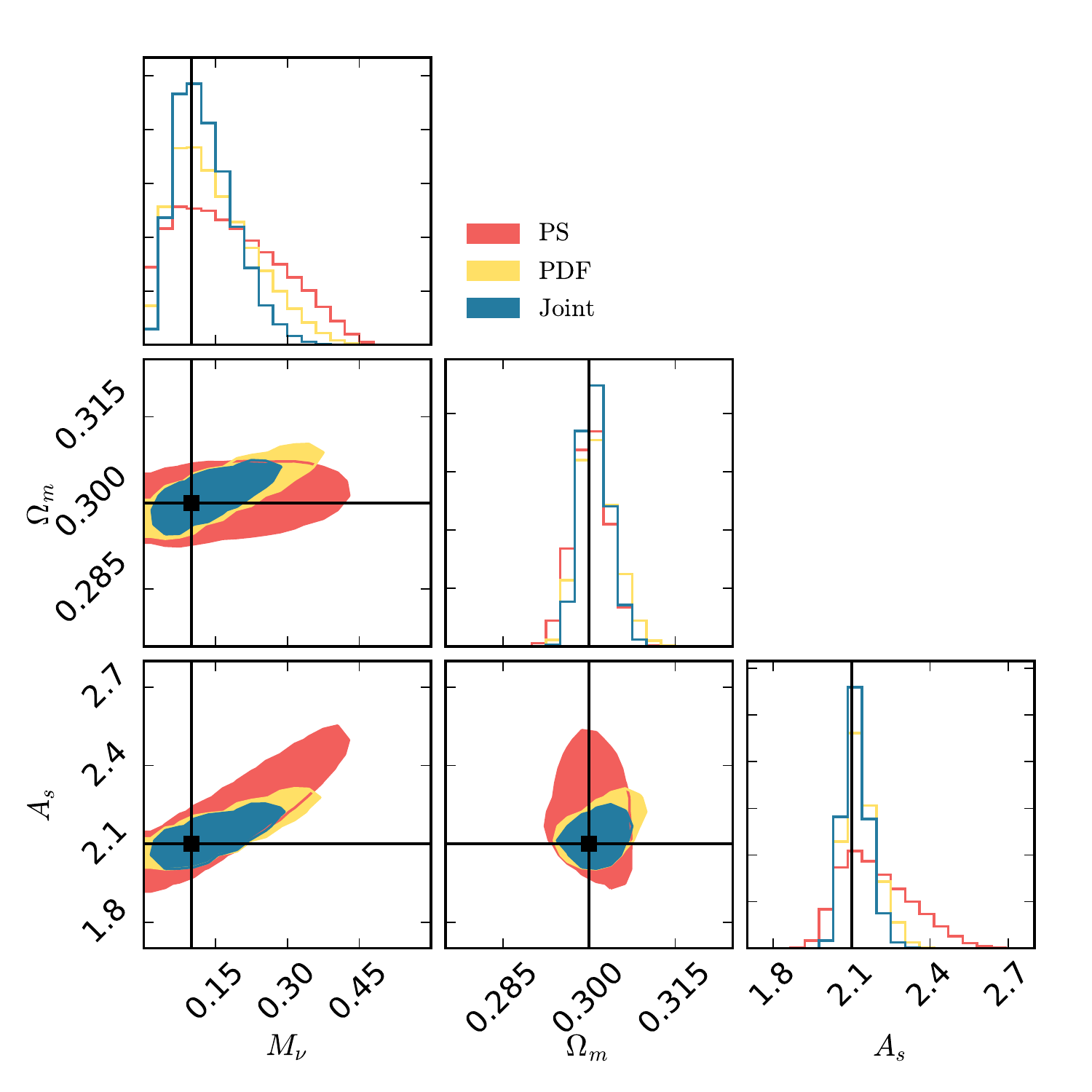}
\end{center}
\caption{\label{fig:joint} 95\% CL contours for weak lensing tomography, for power spectrum~(PS), PDF, and the two jointly. Full covariance is used. We assume an LSST-like survey. }
\end{figure*}

We examine the constraints from combining the power spectrum and PDF. The 95\% CL contours are shown in Fig.~\ref{fig:joint}, all using five tomographic redshift bins. One striking observation is that the PDF along can already outperform the power spectrum. The degeneracy direction of the PDF contour is slightly misaligned with that of the power spectrum. As the result, when joining the two statistics, the combined contour is further shrunk from that of either statistic alone.


We quantify the improved constraints by comparing the 95\% CL PDF and joint errors to that from the power spectrum:
\begin{center}
\begin{tabular}{c|c|c|c} 
 \multicolumn{4}{r}{$\sigma_p/\sigma_p^{\rm power\,spectrum}$}\\
\hline
$p$	&	$\Sigma m_\nu$ & $\Omega_m$ & $A_s$	\\
\hline
power spectrum &	1.00	&	1.00	&	1.00\\
\hline
PDF 			&	0.81	&	1.02	&	0.48\\
\hline
joint 			&	0.65	&	0.84	&	0.39\\
\hline
\end{tabular}
\end{center}
The PDF is particularly powerful in constraining $A_s$, likely due to its sensitivity to a higher power of $A_s$ than the power spectrum. For $\Sigma m_\nu$, the PDF alone is better than the power spectrum by 20\%, and when the two are combined, the error is shrunk by 35\%.

\section{Conclusion}\label{sec:conclusion}

In this paper, we study the constraints from weak lensing tomography on the neutrino mass sum $\Sigma m_\nu$, as well as $\Omega_m$ and $A_s$. We use N-body ray-tracing mocks from the \texttt{MassiveNuS} simulations to fully capture the nonlinear growth in a massive neutrino cosmology. In particular, we attempt to extract additional information beyond the power spectrum, using the one-point PDF. Our main findings are:

(1) Nonlinear growth generates non-Gaussianity in the PDF, demonstrating additional information beyond the power spectrum;

(2) Massive neutrinos suppress both the high and low tails of the PDF, likely the result of reduced number of massive halos and troughs. The suppression is sensitive to the source redshift;

(3) Tomography helps tighten the constraints for both the power spectrum and PDF, by 20--60\% for the parameters studied, when compared to using one single redshift bin; and

(4) The weak lensing PDF alone outperforms the power spectrum in constraining cosmology, consistent with findings by Ref.~\cite{Patton2017}. When the two statistics are combined, the constraints are further tightened by 35\%, 15\%, 61\% for $\Sigma m_\nu$, $\Omega_m$, $A_s$, respectively, when compared to using the power spectrum alone.

In summary, tomographic measurements of the PDF of galaxy weak lensing convergence can help us access the non-Gaussian information in weak lensing data, and will be powerful in constraining the neutrino mass sum. Here we examine the simple case where only the galaxy shape noise is considered. To realize its full potential in next generation deep/wide galaxy surveys, we need to study the measurement and physical systematics, including multiplicative bias in galaxy shapes, photometric redshift errors,  intrinsic alignments, magnification bias, and baryonic effects. These systematics will likely impact both the power spectrum and PDF. However, the hope is that the effects will be different and may be mitigated using the joint analysis. We defer this question to future work. Finally, we also anticipate that the inclusion of primary CMB and BAO data will significantly help breaking the degeneracy with $A_s$ and $\Omega_m$ and hence further tighten the constraint on $\Sigma m_{\nu}$. 

\begin{acknowledgments}
We thank Will Coulton, Zoltan Haiman, Colin Hill, Zack Li, Gabriela Marques, David Spergel, Francisco Villaescusa-Navarro, Ben Wandelt, 
Jose Manuel Zorrilla for helpful discussions. 
This work is supported by an NSF Astronomy and Astrophysics Postdoctoral Fellowship (to JL) under award AST-1602663. We acknowledge support from the WFIRST project.
We thank New Mexico State University (USA) and Instituto de Astrofisica de Andalucia CSIC (Spain) for hosting the Skies \& Universes site for cosmological simulation products.
This work used the Extreme Science and Engineering Discovery Environment (XSEDE), which is supported by NSF grant ACI-1053575. The analysis is in part performed at the TIGRESS high performance computer center at Princeton University. 
\end{acknowledgments}

\bibliographystyle{physrev}
\bibliography{paper}
\end{document}